\documentstyle[aps,epsf,preprint]{revtex}
\begin{document}
\begin{titlepage}
\pagestyle{empty}
\vspace{1.0in}
\begin{flushright}
October 1997
\end{flushright}
\vspace{.1in}
\begin{center}
\begin{large}
{\bf p-Brane cosmology and phases of Brans-Dicke theory with matter}
\end{large}
\vskip 0.5in
Chanyong Park\footnote{chanyong@hepth.hanyang.ac.kr} and 
Sang-Jin Sin\footnote{sjs@dirac.hanyang.ac.kr}\\
\vskip 0.2in
{\small {\it Department of Physics, Hanyang University\\
Seoul, Korea}}
\end{center}
\vspace{1 cm}

\begin{abstract}

\end{abstract}   
We study the effect of the solitonic degrees of freedom in string cosmology
following the line of Rama.
The gas of  solitonic  p-brane is treated as a perfect fluid 
in a Brans-Dicke type theory. 
In this paper, we find exact cosmological solutions 
for any Brans-Dicke parameter $\omega$ and for general parameter $\gamma$ of 
equation of state and classify the cosmology of the solutions on a parameter 
space of $\gamma$ and $\omega$. 

\vspace{.3in}\noindent PACS numbers:98.80.Cq, 11.25.-w, 04.50.+h
\end{titlepage}

\section{ Introduction}
Recent developments of the string theory suggest that in a 
regime  of Planck length curvature, quantum fluctuation is very large  
so that string coupling becomes large and consequently 
the fundamental string degrees of freedom
are not a weakly coupled {\it good} ones \cite{witten}.
Instead, solitonic degrees of freedom like p-brane or 
D-p-branes \cite{pol} are more important. 
Therefore it is a very interesting question 
to ask what is the effect of these new degrees of freedom to 
the space time structure especially whether including these degrees 
of freedom resolve the initial singularity, which is a problem in 
standard general relativity.

For the investigation of  the p-brane cosmology, 
the usual low energy effective action coming from the beta 
function of the string worldsheet would not a good starting point.
So there will be a difference from the string cosmology \cite{vene}. 
We need to  find the low energy effective theory that contains gravity and 
at the same time reveals the effect of these solitonic objects.
But those are not known. Therefore one  can only  guess the answer at this 
moment. 
The minimum amount of the requirement is that it should give a gravity theory 
therefore it must be a generalization of general relativity. The 
Brans-Dicke (BD) theory \cite{brans} is a generic deformation of the general relativity allowing variable gravity coupling. 
In fact low energy theory of the fundamental string 
contains the Brans-Dicke theory with a fine tuned deformation parameter
($\omega$=-1). Moreover Duff and et al. \cite{du} found that the
natural metric that couples to the p-brane is the  Einstein metric
multiplied by certain power of dilaton field.  In terms of this new
metric, the action that gives the p-brane solution  becomes
Brans-Dicke action with definite deformation  parameter $\omega$
depending on p. Using this action, Rama \cite{ra} recently  argued
that the gas of  solitonic  p-brane \cite{du} treated as 
a perfect fluid type matter in a Brans-Dicke theory can resolve the 
initial singularity without any explicit solution. 
In a previous paper \cite{sgl}, we have studied this model and found
several analytic solutions for a few values of parameter with which
the coupled dilaton-graviton system could be decoupled by the simple
``completing the square'' method. In this paper, we give exact
cosmological solutions  for any Brans-Dicke parameter $\omega$ and for
general equation of  state and classify the cosmology of the
solutions according to the range of parameters involved. 

The rest of this paper is organized as follows.
In section II,  we set up the action for the p-brane cosmology. 
In section III, we find an analytic solution for the equation of
motion and  constraint equation for the general case.
In section IV and V, we study the cosmology of the solution and
classify them according to their behavior. In section IV, $t$ as a
function of the dilaton time $\tau$ is studied and the behavior of
the scale factor $a$ with respect to the dilaton time $\tau$ in the
asymptotic region is studied in section V. In section VI, using the
results in section IV and V, we classify the cosmology into several
phases and investigate the behavior of the scale factor $a$ as a
function of the cosmic time $t$.

In section VII, we summarize and conclude with some discussions.

\section{ Construction of the action with the solitonic matter}

We consider the bosonic part of the effective string action and analyse the evolution
of a $D$ dimensional homogeneous isotropic universe with the solitonic matter 
included. The action is given by
\begin{equation}
S = \int d^D x \sqrt{-g} e^{-\phi} \left[ {\cal R} - \omega \partial_{\mu} \phi
\partial^{\mu} \phi \right] + S_m ,
\end{equation}
where $\phi$ is the dilaton field and $S_m$ is the
matter part of the action. In string
theory the BD parameter $\omega$ is fixed as $-1$. In the high
curvature regime, the string coupling is also big and the solitonic
p-brane will be copiously produced since they become light and
dominate the universe in that  regime. Duff et al. \cite{du} have
shown that in terms of metric which couples minimally to
p-brane(p=d-1), the effective action can be written as Brans-Dicke
theory with the BD parameter $\omega$ given by
\begin{equation}
\omega = - \frac{(D-1)(d-2)-d^2}{(D-2)(d-2)-d^2} . 
\end{equation}
In 4-dimension, the BD parameter is given by $\omega = -\frac{4}{3}$
for the $0$-brane  ($p=0$) and $\omega = - \frac{3}{2}$ for the
instanton ($p=-1$). Let's assume that the gas of solitonic p-brane
can be considered as perfect fluid in the Brans-Dicke theory 
with the equation of state $p=\gamma \rho, \gamma <1$. Therefore our
starting point is the equation of the BD theory \cite{Weinberg},
\cite{Veneziano} 
\begin{eqnarray}
{\cal R}_{\mu\nu} - \frac{g_{\mu\nu}}{2} {\cal R} &=&
       \frac{e^{\phi}}{2} T_{\mu\nu} + \omega  \{ \partial_{\mu} \phi
       \partial_{\nu} \phi - \frac{g_{\mu\nu}}{2} (\partial \phi)^2
       \} \nonumber \\ && + \{ -\partial_{\mu} \partial_{\nu} \phi +
       \partial_{\mu} \phi \partial_{\nu} \phi + g_{\mu\nu} {\cal D}^2
       \phi - g_{\mu\nu} (\partial \phi)^2 \} , \nonumber \\ 
&&{\cal R} - 2\omega {\cal D}^2 \phi + \omega (\partial \phi)^2 = 0 
\end{eqnarray}
where $\phi$ is the dilaton and ${\cal D}$ means a covariant
derivative. ${\cal R}$ is the curvature scalar and cosmological metric
is given as the following form 
\begin{equation}
{ds_D}^2 = -\frac{1}{\cal N} dt^2 + e^{2\alpha(t)} \delta_{ij} dx^i dx^j \;\; 
( i,j = 1, 2, \cdots, D-1) ,
\end{equation}
where $e^{\alpha(t)} (=a(t))$ is the scale factor and ${\cal N}$ is
the (constant) lapse  function. Now, we assume that all variables are
the functions of time only. The curvature scalar \cite{Lu} in $D$
dimension is given by
\begin{eqnarray}
{\cal R} &=& g^{00} {\cal R}_{00} + g^{ij} {\cal R}_{ij}, \nonumber \\
g^{00} {\cal R}_{00} &=& \frac{D-1}{\cal N} [ \ddot{\alpha} 
                             + \dot{\alpha}^2 ], \nonumber \\
g^{ij} {\cal R}_{ij} &=& \frac{D-1}{\cal N} [ \ddot{\alpha} 
                             + (D-1) \dot{\alpha}^2 ]
\end{eqnarray} 
where $\dot{\alpha}$ means the time derivative of $\alpha$.

The energy-momentum tensor of the solitonic matter is given by 
\begin{equation} 
T_{\mu\nu} = p g_{\mu\nu} + (p + \rho) U_{\mu} U_{\nu}
\end{equation}            
where $U_{\mu}$ is the fluid velocity. The hydrostatic equilibrium
condition of energy-momentum conservation is
\begin{equation}
\dot{\rho} + (D-1) (p + \rho) \dot{\alpha} = 0 . 
\end{equation} 
Using $p=\gamma \rho$, we get the 
solution 
\begin{equation} 
\rho = \rho_0 e^{-(D-1)(1+\gamma) \alpha}.
\end{equation}
The parameters $\gamma$ and $\omega$ expressed in 
eq.(1) and eq.(8) are free parameters. Our goal is to study how the
metric variables change their behavior for various values of
$\gamma$ and $\omega$.

If we consider only the time dependence, the action can be brought to
the following form
\begin{eqnarray}
S &=& \int dt \;\; e^{(D-1)\alpha-\phi} \left[ \frac{1}{\sqrt{\cal N}} \big{\{}
      -(D-2)(D-1)\dot{\alpha}^2 + 2 (D-1)\dot{\alpha}\dot{\phi} + \omega\dot{\phi}^2 
      \big{\}} \right. \nonumber \\      
  && \left. \hspace{2.5cm} - {\sqrt{\cal N}} \rho_0 e^{-(D-1)(1+\gamma)\alpha + \phi}
     \frac{}{} \right]. 
\end{eqnarray}
where we eliminated $p$ and $\rho$ by eq.(7) and eq.(8).
The variation over the constant lapse function, which is only
$g_{00}$, gives a constraint equation. When we set the lapse function
${\cal N}$ to be 1 after varying of the action over the lapse function
${\cal N}$, this constraint equation is the equation of motion of the
$g_{00}$ component in eq.(5).

\section{ Analytic solution}

Now, we introduce  a new time variable $\tau$ by 
\begin{equation}
dt = e^{(D-1)\alpha - \phi} d\tau .
\end{equation}
Then the action becomes                                                            
\begin{equation}
S= \int d\tau \left[ \frac{1}{\sqrt{\cal N}} \big{\{} (D-1) \kappa \dot{Y}^2 + \mu
     \dot{X}^2 \big{\}} - \sqrt{\cal N} \rho_0 e^{-2X} \right] ,
\end{equation}
where new variables presented in the action are given by
\begin{eqnarray}
 \kappa &=& (D-1)(1-\gamma)^2 (\omega-\omega_{\kappa}) , \nonumber \\
 \nu    &=& 2(1-\gamma)(\omega-\omega_{\nu}) , \nonumber \\
 \mu    &=& - \frac{4(D-2)}{\kappa} (\omega-\omega_{-1}) , \nonumber \\
 -2X    &=& (D-1)(1-\gamma) \alpha - \phi , \nonumber \\
 Y      &=& \alpha + \frac{\nu}{\kappa} X ,\nonumber \\ 
 \omega_{\kappa} &=& -\frac{D-2D\gamma+2\gamma}{(D-1)(1-\gamma)^2}, \nonumber \\
 \omega_{\nu}    &=& -\frac{1}{1-\gamma}, \nonumber \\
 \omega_{-1}     &=& -\frac{D-1}{D-2}.
\end{eqnarray}
The constraint equation is written as
\begin{equation}
0 = (D-1)\kappa \dot{Y}^2 + \mu \dot{X}^2 + \rho_0 e^{-2X} ,   
\end{equation}
where $\rho_0$ is a positive real constant. The equations of
motion are written as
\begin{eqnarray}
0 &=& \ddot{Y} , \nonumber \\
0 &=& \ddot{X} - \frac{\rho_0}{\mu} e^{-2X} .
\end{eqnarray}
Note that $\omega_{-1}$ in eq.(12) happens to be the value of the
instanton. If $\omega$ is less than $\omega_{-1}$, the kinetic term
of the dilaton has a negative energy in Einstein frame. So we will
consider the case where $\omega$ is larger than $\omega_{-1}$.
According to the sign of $\kappa$, the types of solutions are very
different.

When $\kappa$ is negative, an exact solution becomes

\begin{eqnarray} 
X &=& \ln \big[ \frac{q}{c} \cosh(c\tau) \big], \nonumber \\  
Y &=& A \tau + B ,
\end{eqnarray}
where $c$, $A$, $B$ and $q = \sqrt{\frac{\rho_0}{\mid \mu \mid}}$ are
arbitrary real constants. Using the constraint equation, we determine
$A$ in terms of other variables 
\begin{equation}
A = \frac{c}{\delta}, \hspace{1cm} \rm{with} \hspace{0.5cm} 
\delta=\sqrt{ - \frac{(D-1)\kappa}{\mu}} = \frac{\mid \kappa \mid}{2\sqrt{1+
\omega \frac{D-2}{D-1}}} . 
\end{equation}
If $\kappa$ is zero, then we can obtain a solution of the equations of
motion, but it  does not satisfy the constraint equation. If $\kappa$
is positive, the solution is
\begin{eqnarray}
 X &=& \ln \big[ \frac{q}{c} \mid \sinh(c\tau) \mid \big], \nonumber \\
 Y &=& \frac{c}{\delta} \tau + B.
\end{eqnarray}

\section{ Cosmology of the solution}  

Now, we investigate the relation between the cosmic time $t$ and the
dilaton time $\tau$. Since the solutions of the equations of motion
have different forms, we study the behavior of $t$ as a function of
$\tau$ case by case.

\subsection{ $\kappa < 0$ case}

In this region, $\omega < \omega_{\kappa}$. $\gamma$ is always less
than $1$.  We find the relation  between $t$ and $\tau$ using eq.(10) 
\begin{eqnarray}
t-t_0 = {\int_{\tau_0}}^{\tau} d\tau^{\prime} \;\; \exp &&\left[ \frac{(D-1)\gamma c}
        {\delta} \tau^{\prime} - (2+\frac{(D-1)\gamma \nu}{\kappa}) 
        \ln \{\frac{q}{c} \cosh(c\tau^{\prime}) \} \right. \nonumber \\ 
        &&     \left. + (D-1) \gamma B  \frac{}{} \right] ,
\end{eqnarray}
where $(D-1) \gamma B$ is a constant. This constant can be
ignored in the limit $\tau \rightarrow \pm \infty$. Because
$\frac{dt}{d\tau}$ is always positive definite, $t$ is a monotonic
function of $\tau$. The behavior of $a(t)$ as a
function of $t$ depends crucially on the relation between $t$ and
$\tau$.  When $\tau$ goes to $\pm \infty$, $t$ is reduced to
\begin{equation}
t-t_0 \approx \frac{1}{T_{\pm}} \left( e^{T_{\pm} \tau} - e^{T_{\pm}
              \tau_0} \right),
\end{equation}
where    
\begin{eqnarray}
T_{\pm} &=& \frac{2c}{\mid \kappa \mid} \left[ (D-1)\gamma \sqrt{1+\omega
            \frac{D-2}{D-1}} \pm \{\kappa+(D-1) \gamma (1+\omega(1-\gamma))\}
            \frac{}{} \right]. 
\end{eqnarray}                                   

We define a new concept for our purpose: $t$ is super-monotonic
function of $\tau$ if it is monotonic and $t$ runs entire real line
when $\tau$ does so. When $t$ is super-monotonic function of $\tau$,
the universe evolves from infinite past to infinite future. Otherwise
the scale factor $a(t)$ has a starting (ending) point at a finite
cosmic time $t_i$ ($t_f$) which corresponds to initial (final)
singularity.   As a mapping, $t$ maps the real line
of $\tau$ to
\begin{eqnarray*}
(-\infty, \infty) \;\; & \rm{if} \;\; & T_- < 0 < T_+ , \\ 
(-\infty, t_f ) \;\; & \rm{if} \;\; & T_- < 0 \; \rm{and} \; T_+ <0 , \\ 
(t_i , \infty) \;\; & \rm{if} \;\; & T_- >0 \; \rm{and} \; T_+ >0 ,
\\ (t_i , t_f )   \;\; & \rm{if} \;\; & T_+ < 0 < T_- . 
\end{eqnarray*}

In the limit $\tau \rightarrow \pm \infty$, the condition $T_{\pm} < 0$ is
expressed as
\begin{equation}
(D-1)  \gamma  \sqrt{1+ \omega\frac{D-2}{D-1}} < \mp 
        [ \kappa +(D-1)\gamma \{ 1+\omega (1-\gamma)\} ].
\end{equation} 
This inequality is divided into two cases according to the sign of
$\gamma$. In each case we obtain the different region of $\omega$
satisfying the condition $T_{\pm} < 0$. 
                                    
\subsubsection{  $\gamma > 0$ case }

Because we  have considered  only the case  $\omega >  \omega_{-1}$,
so the  left hand side in  eq.(21) is positive definite. To satisfy the
inequality $T_- <0$, the conditions
\begin{eqnarray}
&& \kappa +(D-1)\gamma \{ 1+\omega (1-\gamma)\} > 0 \;\; \rm{and}
\nonumber \\ &&  \left( (D-1) \gamma \sqrt{1+ \omega \frac{D-2}{D-1}}
\right)^2   <  \left( \kappa +(D-1)\gamma \{ 1+\omega (1-\gamma)\}
\right)^2, 
\end{eqnarray}
must be satisfied. If they don't, we know that
$T_-$ is positive. The first inequality in eq.(22) is reduced to the
following inequality
\begin{equation}
\omega > \omega_{-\infty} := -\frac{D-(D-1)\gamma}{(D-1)(1-\gamma)} .
\end{equation}
It is remarkable that the second inequality in eq.(22) is written as 
\begin{equation} 
(\omega - \omega_0)(\omega-\omega_{\kappa}) > 0 ,
\end{equation}
where $\omega_{\kappa}$ appeared in the definition of $\kappa$ and 
\begin{equation}
\omega_0 = -\frac{D}{D-1}
\end{equation}
is the value of $\omega$ for the 0-brane. See eq.(2). \\

Figure 1.  \\

As shown in Figure 1, $\omega_{\kappa} < \omega_0$ ($\omega_{\kappa} >
\omega_0$) in the region $0 < \gamma < \frac{2}{D}$ ($ \gamma >
\frac{2}{D}$). Therefore the solution of eq.(24) becomes 
\begin{eqnarray} 
\omega_{-1} < \omega < \omega_{\kappa} \;\; &\rm{for}& \;\; 0 < \gamma < \frac{2}{D},
\nonumber \\
\omega_{-1} < \omega < \omega_0 \;\; &\rm{for}& \;\; \gamma > \frac{2}{D}.   
\end{eqnarray}
Combining eq.(23) and eq.(26), we find the region of $\omega$ satisfying the
condition $T_- <0$ as the following
\begin{eqnarray}
\omega_{-1} < \omega < \omega_{\kappa} \;\; &\rm{for}& \;\; \frac{1}{D-1} < \gamma <
\frac{2}{D}, \nonumber \\  
\omega_{-1} < \omega < \omega_0 \;\; &\rm{for}& \;\; \gamma > \frac{2}{D}. 
\end{eqnarray}  

The condition $T_+ < 0$ is  
\begin{equation}
(D-1) \gamma \sqrt{1+ \omega\frac{D-2}{D-1}} < - [ \kappa + (D-1)\gamma
                    \{ 1+ \omega (1-\gamma) \} ]. 
\end{equation}
For this, two conditions 
\begin{eqnarray}
&& \kappa +(D-1)\gamma \{ 1+\omega (1-\gamma)\} < 0 \;\; \rm{and}
\nonumber \\ 
&&\left( (D-1) \gamma  \sqrt{1+ \omega\frac{D-2}{D-1}} \right)^2 <
\left ( - [ \kappa+(D-1)\gamma \{ 1+\omega (1-\gamma)\} ] \right)^2, 
\end{eqnarray}         
must be satisfied at the same time. In eq.(29), 
the first inequality gives 
\begin{equation}
\omega < \omega_{-\infty}
\end{equation}
and the second inequality gives eq.(24) again. Therefore using eq.(26) and
eq.(30), we find the region of $\omega$  satisfying  $T_+ < 0$
\begin{equation}
\omega_{-1} < \omega < \omega_{\kappa} \;\; \rm{for} \;\; 0 < \gamma < \frac{1}{D-1}. 
\end{equation} 

\subsubsection{ $\gamma < 0$ case }
 
In this case, the condition $T_- <0$ is written as
\begin{equation}
(D-1) \mid \gamma \mid \sqrt{1+ \omega\frac{D-2}{D-1}} > - [ \kappa
+(D-1)\gamma \{ 1+\omega (1-\gamma)\} ] .
\end{equation}
For this, we need 
\begin{eqnarray}
&& \kappa +(D-1)\gamma \{ 1+\omega (1-\gamma)\} > 0 \;\; \rm{or} \nonumber \\
&& \left( (D-1) \gamma \sqrt{1+ \omega\frac{D-2}{D-1}} \right)^2  >  \left( 
- [ \kappa +(D-1)\gamma \{ 1+\omega (1-\gamma)\} ] \right)^2. 
\end{eqnarray}
Eq.(33) can be simplified as
\begin{equation}
\omega_{-\infty} <\omega \;\; \rm{or} \;\; \omega_0 < \omega <
\omega_{\kappa} .
\end{equation}
Thus the solution, which is the sum of two regions in eq.(34), is reduced to  
\begin{equation}
\omega_0 < \omega < \omega_{\kappa} \;\; \rm{for} \;\; \gamma <0.
\end{equation}
This solution includes the region of the first
inequality of eq.(34). 

Similarly, the condition $T_+ < 0$ is written as 
\begin{equation}
(D-1) \mid \gamma \mid \sqrt{1+ \omega\frac{D-2}{D-1}} >  \kappa
+(D-1)\gamma \{ 1+\omega (1-\gamma)\},
\end{equation}
which gives
\begin{eqnarray}
&& \kappa +(D-1)\gamma \{ 1+\omega (1-\gamma)\} < 0 \;\; \rm{or} \nonumber \\
&& \left( (D-1) \gamma \sqrt{1+ \omega\frac{D-2}{D-1}} \right)^2  >  \left( 
 \kappa +(D-1)\gamma \{ 1+\omega (1-\gamma)\}  \right)^2. 
\end{eqnarray}
The solution of these can be written as $\omega < \omega_{-\infty}
\; \rm{or} \; \omega_0 < \omega < \omega_{\kappa}$. From these, the
region satisfying $T_+ < 0$ is
\begin{equation}
\omega_{-1} < \omega < \omega_{\kappa} \;\; \rm{for} \;\; \gamma < 0.
\end{equation}

\subsection{ $\kappa > 0$ case}

Now we consider positive $\kappa$, which means 
\begin{equation}
\omega > \omega_{\kappa}.
\end{equation} 
Since the solution $X(\tau)$ has a singularity at
$\tau=0$, we have to treat carefully the behavior of $t$ near
$\tau=0$. The relation between $t$ and $\tau$ is given by 
\begin{eqnarray}
t-t_0 = {\int_{\tau_0}}^{\tau} d\tau^{\prime} \;\; \exp &&\left[ \frac{(D-1)\gamma c}
        {\delta} \tau^{\prime} - (2+\frac{(D-1)\gamma \nu}{\kappa}) 
        \ln \{\frac{q}{c} \mid \sinh(c\tau^{\prime}) \mid \} \right. \nonumber \\
        &&     \left. + (D-1) \gamma B  \frac{}{} \right] .
\end{eqnarray}
In the limit $\tau \rightarrow 0$, the above equation is reduced to 
\begin{equation}
t - t_0 \approx \rm{sign}(\tau) \frac{q^{-\eta} e^{(D-1)\gamma
      B}}{1-\eta} \left[ \mid \tau \mid^{1-\eta} - \mid \tau_0
      \mid^{1-\eta} \right],
\end{equation} 
where $\eta = 2+\frac{(D-1)\gamma \nu}{\kappa}$ and  $\tau_0$ and
$t_0$ are real  constants. In case of $\eta > 1$, $t$ has a
singularity at $\tau \rightarrow 0$. In the other case, $t$ has no
singularity. So we consider two cases $\eta <1$ and $\eta >1$.

\subsubsection{ $\eta <1$ case}

In this case, $t$ has no singularity at $\tau =0$. So we
investigate the behavior of $t$ at $\tau \rightarrow \pm \infty$ only.

\noindent i) $\gamma > 0$ case  

In the case $\kappa >0$, eq.(40) is reduced to
\begin{equation}
t-t_0 \approx \frac{1}{T_{\pm}} \left( e^{T_{\pm}  \tau} -
                 e^{T_{\pm} \tau_0} \right), 
\end{equation}
where
\begin{equation}
T_{\pm} = \frac{2c}{\mid \kappa \mid} \left[ (D-1)\gamma \sqrt{1+\omega
            \frac{D-2}{D-1}} \mp \{\kappa+(D-1) \gamma (1+\omega(1-\gamma))\}
            \frac{}{} \right]. 
\end{equation}
The condition $T_- < 0$ is written as
eq.(28) and gives the solution
\begin{eqnarray}
&& \omega < \omega_{-\infty} \; \rm{and} \; \omega > \omega_0 \;\;
\rm{for} \;\; 0 < \gamma < \frac{2}{D}, \nonumber \\ 
&& \omega < \omega_{-\infty} \; \rm{and} \; \omega >
\omega_{\kappa} \;\; \rm{for} \;\; \gamma > \frac{2}{D},
\end{eqnarray}
where we use $\omega > \omega_{\kappa}$. As shown Figure 1, 
$\omega_0 > \omega_{-\infty}$ for $0 < \gamma < \frac{2}{D}$
and  $\omega_{\kappa}  > \omega_{-\infty}$ for $\gamma  >
\frac{2}{D}$. Therefore there is no solution satisfying the  condition 
$T_- <0$. Hence $T_-$ is positive.

Now we investigate the behavior of $t$ at $\tau \rightarrow
+\infty$. The condition $T_+ < 0$ is written like eq.(22). Applying
the similar method used in the above analysis, the region of $\omega$
satisfying $T_+ < 0$ is summarized as the following
\begin{eqnarray}
\omega_0 < \omega \;\; &\rm{for}& \;\; 0 < \gamma < \frac{2}{D}, \nonumber \\
\omega_{\kappa} < \omega \;\; &\rm{for}& \;\; \gamma > \frac{2}{D}.
\end{eqnarray}

\noindent ii) $\gamma < 0$ case 
                 
Through the same calculation, we can show that $T_-$ is
positive and $T_+$ is negative for all negative $\gamma$. 

\subsubsection{ $\eta > 1$ case }

In this case, the behavior of $t$ is singular at $\tau = 0$.
$\tau_0$ and $t_0$ can be ignored due to the divergence of $\mid
\tau \mid^{1-\eta}$. From eq.(40) or eq.(41), we know that
$\frac{dt}{d\tau}$ is always positive definite  except a singular
point $\tau=0$. The condition $\eta > 1$ is reduced to 
\begin{equation}
\omega > -\frac{D}{(D-1)(1-\gamma^2)}  := \omega_{\eta}  .
\end{equation}
Under this condition, the region of $\tau$ is divided into $-\infty <
\tau < 0$ and $0< \tau < \infty$. Near $\tau = 0$, we obtain the
behavior of $t$ characterized by the sign of $\tau$. When $\tau$ goes
to zero from below, $t$ in eq.(41) is written as  
\begin{equation}
t \approx \frac{q^{-\eta} e^{(D-1) \gamma B}}{(\eta-1)}
\frac{1}{(-\tau)^{\eta-1}}.
\end{equation}
When $\tau$ goes to zero from above, $t$ is reduced as the following
\begin{equation} 
t \approx -\frac{q^{-\eta} e^{(D-1) \gamma B}}{(\eta-1)}
\frac{1}{\tau^{\eta-1}}.
\end{equation}
Thus $t \rightarrow +\infty$ as $\tau \rightarrow -0$ but $t
\rightarrow -\infty$ as  $\tau \rightarrow +0$. We also have to
examine the behaviors of $t$ at $\tau \rightarrow \pm \infty$.
However these have been already described when we discussed
the case $\eta <1$. \\

To describe the behavior of $t$ as a function of $\tau$ at $\tau
\rightarrow \pm \infty$ and $\tau \rightarrow 0$, we classify
the parameter space of $\gamma$ and $\omega$ using all results
obtained in this section. These are shown in Figure 2. \\

Figure 2 \\

The behavior of the $t$ as a function of $\tau$ in Figure 2 is
summarized as the followings:  

1) In region I, $T_- > 0$ and $T_+ <0$ ($\omega < \omega_{\kappa}$).
\\  $t$ evolves from finite initial time $t_i$ to finite final
time $t_f$ as $\tau$ runs ($-\infty, +\infty$).

2) In region II, $T_- < 0$ and $T_+ <0$ ($\omega < \omega_{\kappa}$).
\\ $t$ evolves from negative infinity to
finite final time $t_f$ as $\tau$ runs ($-\infty, +\infty$).

3) In region III, $T_- > 0$ and $T_+ <0$ ($\omega > \omega_{\kappa}$
and $\omega > \omega_{\eta}$). \\
In this region, because $t$ has  a singular behavior at $\tau=0$, the
region of $\tau$ divided  into $-\infty < \tau < 0$ and  $0 < \tau
<\infty$. Therefore $t$ has two branches for any given values of
$\gamma$ and $\omega$. For $-\infty < \tau < 0$, $t$ evolves from
finite initial time $t_i$ to positive infinity.  For $0 < \tau
<\infty$, $t$ evolves  from negative infinity to finite final time
$t_f$.

4) In region IV, $T_- > 0$ and $T_+ >0$ ($\omega > \omega_{\kappa}$
and $\omega < \omega_{\eta}$). \\
In this region, because $t$  has no singularity, $t$ evolves from
finite initial time  $t_i$  to positive infinity as $\tau$ runs
($-\infty, +\infty$).

5) In region V, $T_- > 0$ and $T_+ >0$ ($\omega > \omega_{\kappa}$
and $\omega > \omega_{\eta}$). \\
By the same reason  that explained in region III, $t$  evolves from
finite initial time $t_i$  to positive infinity for $-\infty <  \tau
< 0$ and  $t$ evolves from negative infinity to positive infinity for
$0  < \tau  <\infty$.

6) In region VI, $T_- < 0$ and $T_+ >0$ ($\omega < \omega_{\kappa}$). \\ 
$t$ evolves from negative infinity to
positive infinity as $\tau$ runs ($-\infty, +\infty$).

7) In region VII, $T_- > 0$ and $T_+ >0$ ($\omega < \omega_{\kappa}$). \\
$t$ evolves from finite initial time $t_i$ to positive infinity
as $\tau$ runs ($-\infty, +\infty$).

\section{ The behavior of the scale factor}

Now we study the behavior of the scale factor $a$ as a function
of $\tau$.

\subsection{ $\kappa < 0$ case}  

We consider the exponent of the scale factor
$\alpha(\tau)$. Using eq.(10), $\alpha(\tau)$ is given by 
\begin{equation}
\alpha(\tau) = \frac{2}{\mid \kappa \mid} 
   \left[ \sqrt{1 + \omega \frac{D-2}{D-1}} c \tau 
   +\{ 1 + \omega (1-\gamma) \} \ln \{ \frac{q}{c} \cosh (c\tau) \} \right] +B .
\end{equation}  
In the limit $\tau \rightarrow \pm \infty$, the scale factor $a(\tau) =
e^{\alpha(\tau)}$ is rewritten as the following
\begin{equation}
a(\tau) \approx e^{H_{\pm} \tau},
\end{equation}
where $H_{\pm}$ is defined as 
\begin{equation}
H_{\pm} =  \frac{2c}{\mid \kappa \mid} \left[ \sqrt{1 + \omega \frac{D-2}{D-1}} 
           \pm \{ 1 + \omega (1-\gamma) \} \right] .
\end{equation} 
Just as $t(\tau)$, the behavior of $a(\tau)$ at $\tau \rightarrow \pm
\infty$ is determined by the sign of $H_{\pm}$. Using this and the
the sign of $T_{\pm}$, we can read the behavior
of the scale factor $a(t)$ as a function of $t$ in the asymptotic
regions.

For negative $\kappa$ ($\omega < \omega_{\kappa}$), $H_- >0$ can be
written as
\begin{equation}  
\sqrt{1+\omega \frac{D-2}{D-1}} > 1 + \omega (1-\gamma). 
\end{equation}
This means
\begin{eqnarray}
&&1 + \omega (1-\gamma) < 0 \;\; \rm{or} \nonumber \\
&& \left( \sqrt{1+\omega \frac{D-2}{D-1}} \right)^2 >
\left( 1 + \omega (1-\gamma) \right)^2 .
\end{eqnarray}
The first inequality in eq.(53) is equivalent to
\begin{equation}
\omega < \omega_{\nu},
\end{equation}
where $\omega_{\nu}$ has been already written in eq.(12). As one can
see in Figure 3, $\omega_{\kappa} < \omega_{\nu}$ if $\gamma <
\frac{1}{D-1}$ and $\omega_{\kappa} > \omega_{\nu}$ if $\gamma >
\frac{1}{D-1}$. Together with $\omega < \omega_{\kappa}$, the first
inequality condition gives the region of $\omega$ satisfying $H_- >0$
\begin{equation}
\omega_{-1} < \omega < \omega_{\kappa} \;\; \rm{for} \;\; \gamma < \frac{1}{(D-1)}.
\end{equation}
The second inequality in eq.(53) is rewritten as
\begin{equation}
\omega (\omega - \omega_{\kappa}) < 0 .
\end{equation}  
Notice that $\omega_{\kappa} < 0$ if $\gamma < \frac{D}{2(D-1)}$ and 
$\omega_{\kappa} > 0$ if $\gamma > \frac{D}{2(D-1)}$. 
Since $\omega < \omega_{\kappa}$, we can rewrite eq.(56)
as the following
\begin{equation}
0 < \omega < \omega_{\kappa} \;\; \rm{for} \;\; \gamma > \frac{D}{2(D-1)}.
\end{equation} 
As a result, eq.(55) and eq.(57) are the regions of $\omega$
satisfying the condition $H_- >0$. \\

Figure 3 \\

Now we consider the condition $H_+ >0$ 
\begin{equation}  
\sqrt{1+\omega \frac{D-2}{D-1}} > - [1 + \omega (1-\gamma) ]. 
\end{equation}
Like the case $H_- >0$, this inequality is divided into two
inequalities
\begin{eqnarray}
&&1 + \omega (1-\gamma) > 0 \;\; \rm{or}  \nonumber \\
&& \left( \sqrt{1+\omega \frac{D-2}{D-1}} \right)^2 >
\left( 1 + \omega (1-\gamma) \right)^2 .
\end{eqnarray}
The first inequality gives the region of $\omega$ satisfying the
condition $H_+ >0$
\begin{equation}
\omega > \omega_{\nu}.
\end{equation}
Using $\omega > \omega_{-1}$ and $\kappa < 0$, eq.(60) is rewritten as
the following
\begin{equation}
\omega_{-1} < \omega < \omega_{\kappa} \;\; \rm{for} \;\; \gamma >
\frac{1}{D-1}.
\end{equation}
The second inequality in eq.(59) has the same region of
$\omega$ that appeared in eq.(57). Because the region of $\omega$ in 
eq.(61) contains the region of $\omega$ in eq.(57), eq.(61) is the
solution satisfying the condition $H_+ >0$.

\subsection{ $\kappa > 0$ case}

The exponent of the scale factor $\alpha(\tau)$ is given by
\begin{equation}
\alpha(\tau) = \frac{2}{\mid \kappa \mid} 
\left[ \sqrt{1 + \omega \frac{D-2}{D-1}} c \tau - \{ 1 + \omega 
(1-\gamma) \} \ln \{ \frac{q}{c} \mid \sinh (c\tau) \mid \} \right] +B .
\end{equation}

\subsubsection{ $\eta < 1$ case }
            
In this case, the scale factor $a(\tau)$ has no singular
behavior. So we investigate the behavior of $a$ at $\tau
\rightarrow \pm \infty$.

In the limit $\tau \rightarrow \pm \infty$, $a(\tau)$ is given by
\begin{equation}
a(\tau) \approx e^{H_{\pm} \tau},
\end{equation}
where $H_{\pm}$ is defined as
\[
H_{\pm} =  \frac{2c}{\mid \kappa \mid} \left[ \sqrt{1 + \omega
   \frac{D-2}{D-1}} \mp \{ 1 + \omega (1-\gamma) \} \right] . 
\] 

The condition $H_- >0$ is exactly equal to eq.(58) due
to the sign of $\kappa$.  When we solve eq.(58) under the condition
$\kappa >0$, $\omega > \omega_{\kappa}$ instead of $\omega <
\omega_{\kappa}$ must be applied to the solution. Then we obtain the
region of $\omega$  satisfying the condition $H_- >0$
\begin{equation}
\omega > \omega_{\kappa} \;\; \rm{for} \;\; \rm{all} \; \gamma.
\end{equation}
The condition $H_+ >0$ described by eq.(52) and  $\omega >
\omega_{\kappa}$ gives the region of $\omega$
\begin{equation}
\omega_{\kappa} < \omega < 0 \;\; \rm{for} \;\; \gamma < \frac{D}{2(D-1)}.
\end{equation}

\subsubsection{ $\eta > 1$ case }

In this case, we need to investigate the behavior of $a(\tau)$ at
$\tau \rightarrow 0$ because $a(\tau)$ has a singular behavior
at $\tau = 0$. In the limit $\tau \rightarrow 0$, $a(\tau)$ is
written as
\begin{equation}
a(\tau) \approx  e^B (q  \mid \tau \mid)^{-\frac{2(1-\gamma)(\omega
           -\omega_{\nu})}{\mid \kappa  \mid}}.
\end{equation}
For $\omega  > \omega_{\nu}$,  where $-(1-\gamma)(\omega  -
\omega_{\nu})$  is  negative, $a(\tau)$ goes to infinite at $\tau
\rightarrow 0$. And for $\omega < \omega_{\nu}$, $a(\tau)$ goes to
zero at $\tau \rightarrow 0$.  The behavior of $a(\tau)$ at $\tau
\rightarrow \pm \infty$ has been described already when we discussed
the case $\eta < 1$. \\

From these studies, we calssify the behavior of $a(\tau)$ on the
parameter space of $\gamma$ and $\omega$. This is shown in Figure 4.
\\

Figure 4. \\

As shown in Figure 4, we summarize the behavior of $a(\tau)$ as the
followings: \\

1) In the region I, $H_- >0$ and $H_+ <0$ ($\omega <
\omega_{\kappa}$). \\ 
In the limit $\tau \rightarrow \pm \infty$,
$a(\tau)$ goes to a zero size. 

2) In the region II, $H_- <0$ and $H_+ >0$ ($\omega <
\omega_{\kappa}$). \\ 
In the limit $\tau \rightarrow \pm \infty$,
$a(\tau)$ goes to an infinite size.

3) In the region III, $H_- >0$ and $H_+ >0$ ($\omega <
\omega_{\kappa}$). \\ 
In the limit $\tau \rightarrow -\infty$, $a(\tau)$ goes to a zero
size. And in the limit  $\tau  \rightarrow \infty$, $a(\tau)$ goes to
an infinite size.

4) In the region IV, $H_- >0$ and $H_+ >0$ ($\omega >
\omega_{\kappa}$ and  $\omega < \omega_{\eta}$). \\
In this region, the behavior of $t$ is not singular. So we need not
to consider the behavior of $a(\tau)$ at  $\tau =0$. In the  limit
$\tau \rightarrow -\infty$, $a(\tau)$  goes to a zero  size.
And in the limit $\tau \rightarrow \infty$, $a(\tau)$ goes to an
infinite size.

5) In  the region V,  $H_- >0$  and $H_+ >0$  ($\omega >
\omega_{\kappa}$,  $\omega >  \omega_{\eta}$ and $\omega <
\omega_{\nu}$). \\ 
In this region, because $t$ has a singular behavior
at $\tau =0$, we interpret the behavior of $a(\tau)$ as the following. 

For $-\infty < \tau < 0$, $a(\tau)$ goes to a zero size at
$\tau \rightarrow  -\infty$ and $\tau \rightarrow 0$. For $0 <
\tau < \infty$, $a(\tau)$ goes to a zero size at $\tau \rightarrow 0$
and goes to an infinite size at $\tau  \rightarrow \infty$.

6) In  the region VI,  $H_- >0$ and $H_+  >0$ ($\omega >
\omega_{\kappa}$, $\omega >  \omega_{\eta}$
and $\omega > \omega_{\nu}$). \\
For $-\infty < \tau < 0$, $a(\tau)$ goes to a zero size at
$\tau \rightarrow  -\infty$ and goes to an infinite size at $\tau
\rightarrow 0$. For $0 < \tau < \infty$, $a(\tau)$ goes to an
infinite size at $\tau \rightarrow 0$ and at $\tau \rightarrow
\infty$.
                                             
7) In  the region VII, $H_-  >0$ and $H_+ <0$  ($\omega >
\omega_{\kappa}$, $\omega  >  \omega_{\eta}$
and $\omega > \omega_{\nu}$). \\
For $-\infty < \tau < 0$, $a(\tau)$ goes to a zero size at
$\tau \rightarrow  -\infty$ and goes to an infinite size at $\tau
\rightarrow 0$. For $0 < \tau < \infty$, $a(\tau)$ goes to an
infinite size at $\tau \rightarrow 0$ and goes to a zero size at $\tau
\rightarrow \infty$. \\

\section{ The phase of cosmology }

Using all results obtained from the section IV and V, we now classify
the parameter space of $\gamma$ and $\omega$ into several phases and
find the behavior of $a(t)$. These phases are characterized according
to the behavior of $a(t)$.  

Using eq.(19) and eq.(50), in the limit $\tau \rightarrow \pm
\infty$, $a(t)$ is written as
\begin{eqnarray}
a(t) \approx [T_- (t- t_i )]^{H_- / T_-} \;\; &\rm{at}& \;\; \tau
\rightarrow -\infty,  \nonumber \\ 
a(t) \approx [T_+ (t- t_f )]^{H_+ / T_+} \;\; &\rm{at}& \;\; \tau
\rightarrow \infty, 
\end{eqnarray}
where $t_i$ and $t_f$, which are defined in section IV, are real
constant. Notice that  $t_i$ ($t_f$) becomes the starting point 
(the ending point) in the case $T_- >0$ ($T_+ <0$) and that $t_i$
($t_f$) can be neglected in the case $T_- <0$ ($T_+ >0$) because $t
\rightarrow \pm \infty$ as $\tau \rightarrow \pm \infty$.

Now, we explain two examples:

i) For $T_- <0$ and $H_- / T_- > 0$, $T_- (t- t_i )$ is positive and
$a(t)$ goes to  positive infinite at $t \rightarrow - \infty$.

ii) For $T_- >0$ and  $H_- / T_- > 0$, $t$ can be defined in the
region $t>t_i$ only. So  $t-t_i$ 
is positive and $a(t)$ goes to zero at $t \rightarrow t_i$.

Other cases can be analysed through the same method.

In the region $\omega > \omega_{\kappa}$ and $\eta > 1$, we must
investigate the behavior  of $a(t)$ at $\tau \rightarrow 0$.  From
eq.(47), eq.(48) and eq.(66), $a(t)$ is obtained as the following
\begin{equation}
a(t) \approx E \times \mid t
\mid^{\frac{2(1-\gamma)(\omega-\omega_{\nu})}{(\eta-1)  \mid \kappa
\mid}}, 
\end{equation}  
where 
\[
E = [q(\eta -1)]^{\frac{2(1-\gamma)(\omega-\omega_{\nu})}{(\eta-1)
\mid \kappa \mid}}  e^{B \left[1-  \frac{2(D-1)\gamma (1-\gamma)
(\omega-\omega_{\nu})}{(\eta-1) \mid  \kappa  \mid} \right] } 
\]
is a positive value because $q$ and $(1-\gamma)$ are positive  in the
previous definition. $a(t)$ goes to zero at $t \rightarrow  \pm
\infty$  ($\tau  \rightarrow  \pm  0$) in the case $\omega <
\omega_{\nu}$ and $a(t)$ goes to infinite at $t \rightarrow \pm
\infty$ in the case $\omega > \omega_{\nu}$. \\

Figure 5 \\

As shown in Figure 5, using the sign of $T_{\pm}$ and $H_{\pm}$ with
the consideration of the behavior of $a(t)$ at $\tau \rightarrow 0$,
the behavior of $a(t)$ in each region is  characterized as the
followings:

1) In region I, $T_- >0$, $T_+ <0$, $H_- >0$, and $H_+ <0$. \\    
The universe evolves from a zero size at finite initial time $t_i$ 
to a zero size at finite final time $t_f$.

2) In region II, $T_- <0$, $T_+ <0$, $H_- >0$, and $H_+ <0$. \\
The universe evolves from a zero size at negative infinity to a zero
size at finite final time $t_f$.

3) In region III, $T_- >0$, $T_+ >0$, $H_- >0$, and $H_+ >0$. \\
The universe evolves from a zero size at finite initial time $t_i$ 
to an infinite size at positive infinity.

4) In region IV, $T_- <0$, $T_+ >0$, $H_- <0$, and $H_+ >0$. \\
The universe evolves from an infinite size at negative infinity
to an infinite  size at positive infinity.

5) In region V, $T_- >0$, $T_+ >0$, $H_- <0$, and $H_+ >0$. \\
The universe evolves from an infinite size at finite initial time
$t_i$  to an infinite size at positive infinity.

6) In region VI, $T_- >0$, $T_+ >0$, $H_- >0$, and $H_+ >0$. \\
The universe evolves from a zero size at finite initial time $t_i$ 
to an infinite size at positive infinity.
                                                                
From 7) to 11), we consider the case $\eta  > 1$ and $\omega >
\omega_{\kappa}$ in which  $t$ has a singular behavior at  $\tau
\rightarrow 0$. In these cases, we can divide  the region of $\tau$
into  $-\infty < \tau <0$ and $0< \tau < \infty$. Therefore we obtain
two branches of $a(t)$ having  the different behaviors in each region
of $\tau$. 

7) In region VII, $T_- >0$, $T_+ >0$, $H_- >0$, and $H_+ >0$. 

In the region $-\infty < \tau <0$, the universe evolves from a zero
size at finite initial time  $t_i$  to a zero size at positive
infinity.

In the region $0< \tau < \infty$, the universe evolves from a zero
size at negative infinity to an infinite size
at positive infinity.

8) In region VIII, $T_- >0$, $T_+ <0$, $H_- >0$, and $H_+ >0$.

In the region $-\infty < \tau <0$, the universe evolves from a zero
size at finite initial time  $t_i$  to a zero size at positive
infinity.

In the region $0< \tau < \infty$, the universe evolves from a zero
size at negative infinity to an infinite size
at finite final time $t_f$.

9) In region IX, $T_- >0$, $T_+ >0$, $H_- >0$, and $H_+ >0$. 

In the region $-\infty < \tau <0$, the universe evolves from a zero
size at finite initial time  $t_i$  to an infinite size at positive
infinity.

In the  region $0<  \tau <  \infty$, the  universe evolves from  an
infinite  size at  negative  infinity to an
infinite size at positive infinity.

10) In region X, $T_- >0$, $T_+ <0$, $H_- >0$, and $H_+ >0$. 

In the region $-\infty < \tau <0$, the universe evolves from a zero
size at finite initial time  $t_i$  to an infinite size at
positive infinity.

In the  region $0<  \tau <  \infty$, the  universe evolves from  an
infinite  size at  negative infinity to
an infinite size at finite final time $t_f$.
 
11) In region XI, $T_- >0$, $T_+ <0$, $H_- >0$, and $H_+ <0$.

In the region $-\infty < \tau <0$, the universe evolves from a zero
size at finite initial time  $t_i$  to an infinite size at
positive infinity.

In the  region $0<  \tau <  \infty$, the  universe evolves from  an
infinite  size at  negative  infinity to a zero size at finite final
time $t_f$.

\section {Discussion and conclusion}

In this paper we studied the effect of the gas of  solitonic  p-brane 
by treating them  as a perfect fluid 
in the Brans-Dicke theory. We found exact cosmological solutions 
for any Brans-Dicke parameter $\omega$ and for general constant
$\gamma$ and classified the cosmology of the solutions
according to  the parameters involved. 
We assumed that the universe is dominated by one kind of p-brane and
they are treated as perfect fluid. We found the analytic solution
which is singularity free  for some $\gamma$ and $\omega$. 
It is very interesting that $a(t)$ has no initial and final
singularities at finite initial and final cosmic time in region IV
and in region VII. $a(t)$ has also an inflation behavior in region
VII. So we need to study more intensively the behavior of $a(t)$ in
these regions.

Presumably the value of $\gamma$ as well as $\omega$ should be fixed
once p is fixed. Without knowing the  value of $\gamma$ for a given
p, the classification was the best thing we could do. It would be
very interesting to determine the parameter $\gamma$ for  the given p.
Also we need more rigorous justification of our basis for the p-brane
cosmology. If what we took as basis goes wrong, then what we have done
is just Brans-Dicke cosmology in the presence of some perfect fluid
type matter. We wish that more study of the effect of the solitons in
the string cosmology be done in the future.

\vskip 1cm
\noindent{\bf \Large Acknowledgement}

\noindent This work has been supported by the research grant KOSEF 971-0201-001-2.

\newpage
 
\vspace{3cm}

Caption \\ \\ \\ \\

Figure 1. In 4 dimension ($D=4$), all functions ($\omega_{\kappa},
\omega_{\eta}, \cdots$), defined by the relation between $t$ and
$\tau$, are presented on a parameter space of $\gamma$ and $\omega$.
\\ \\

Figure 2. In 4 dimension ($D=4$), the parameter space is classified
by the relation between $t$ and $\tau$. \\ \\

Figure 3. In 4 dimension ($D=4$), all functions ($\omega_{\kappa},
\omega_{\nu}, \cdots$), defined by the relation between $a(\tau)$ 
and $\tau$, are presented on a parameter space of $\gamma$ and 
$\omega$. \\ \\

Figure 4. In 4 dimension ($D=4$), the parameter space is classified
by the behavior of $a(\tau)$. \\ \\

Figure 5. In 4 dimension ($D=4$), the parameter space is classified
by the behavior of $a(t)$. \\ \\

\newpage

\begin{figure}
\begin{center}
$${\epsfxsize=10.0 truecm \epsfbox{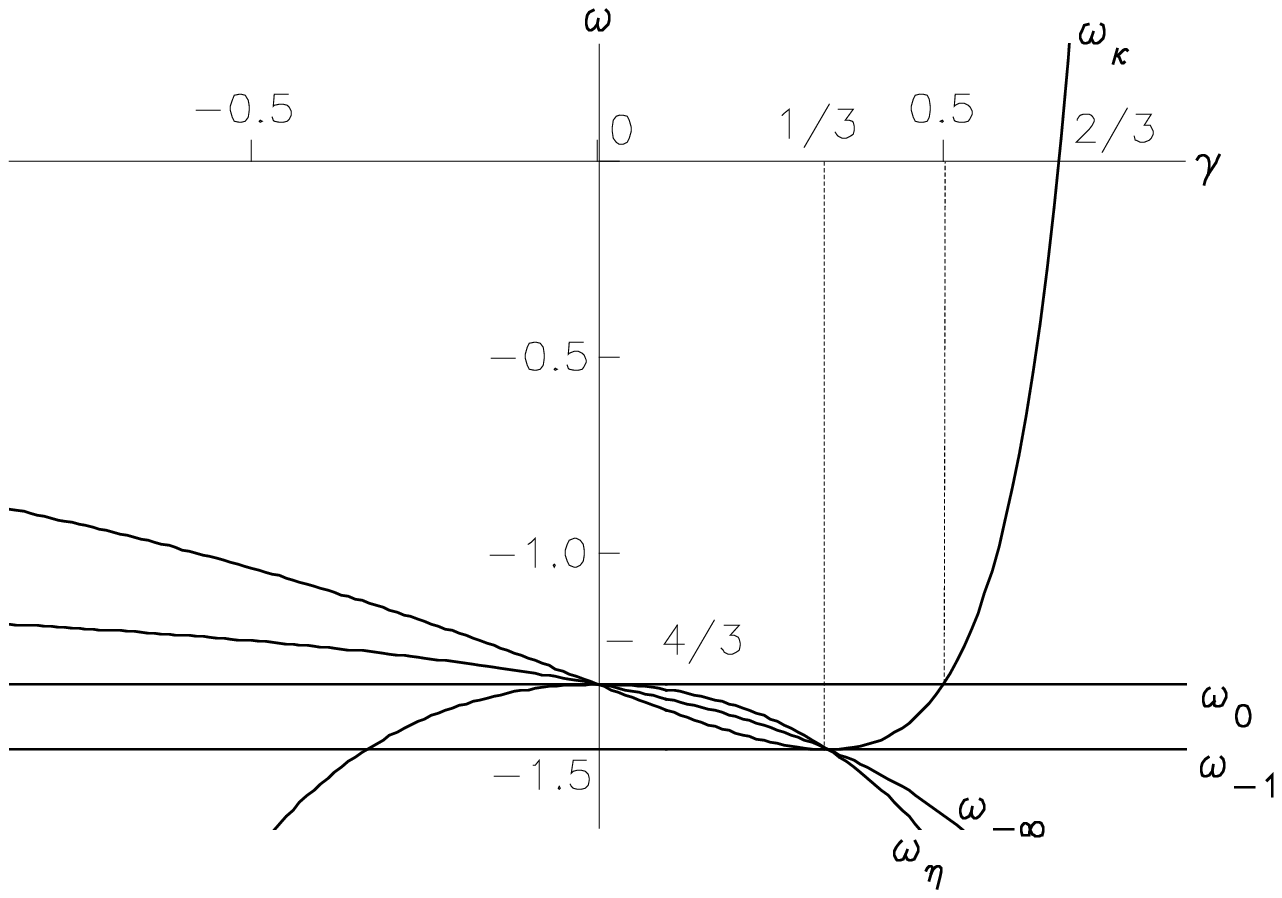}}$$
\end{center}
\end{figure}

\begin{figure}
\begin{center}
$${\epsfxsize=10.0 truecm \epsfbox{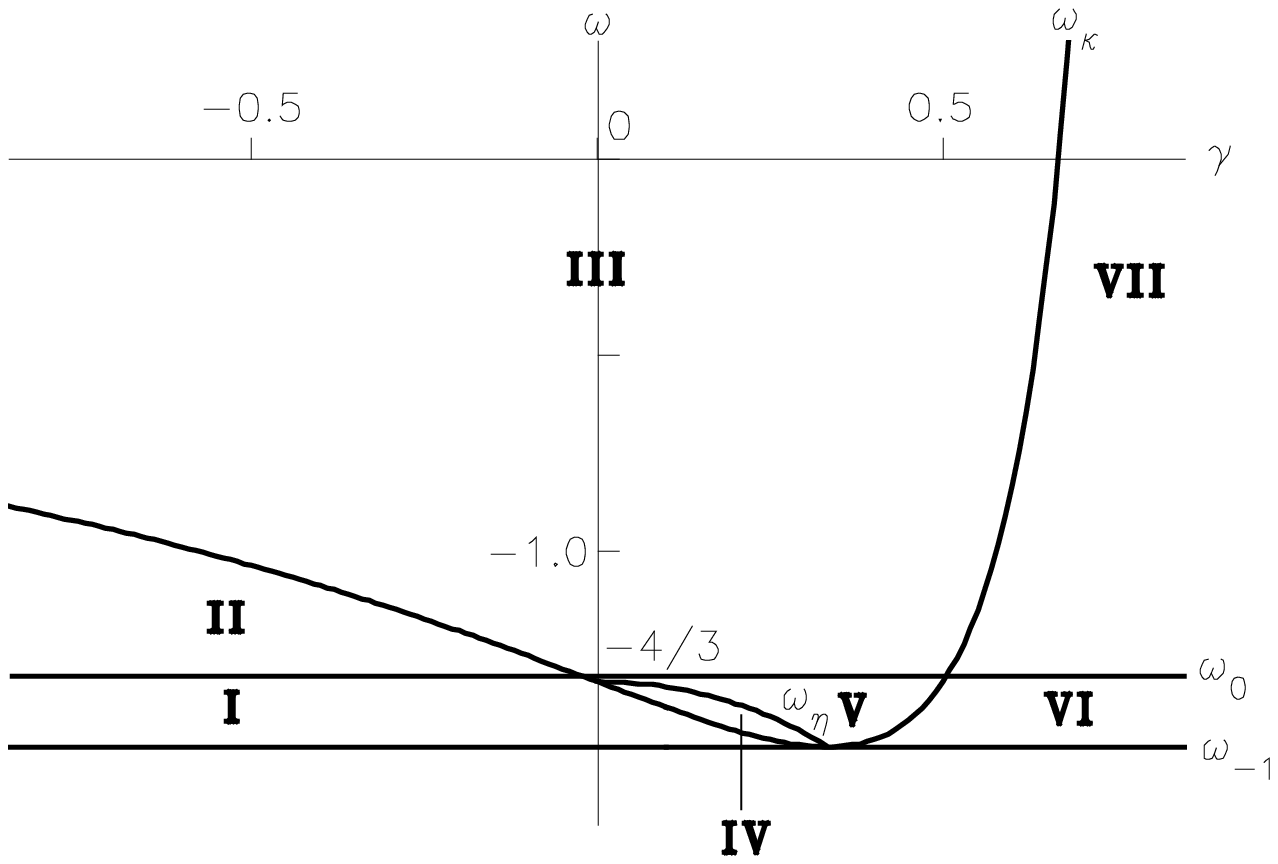}}$$
\end{center}
\end{figure}

\newpage

\begin{figure}
\begin{center}
$${\epsfxsize=10.0 truecm \epsfbox{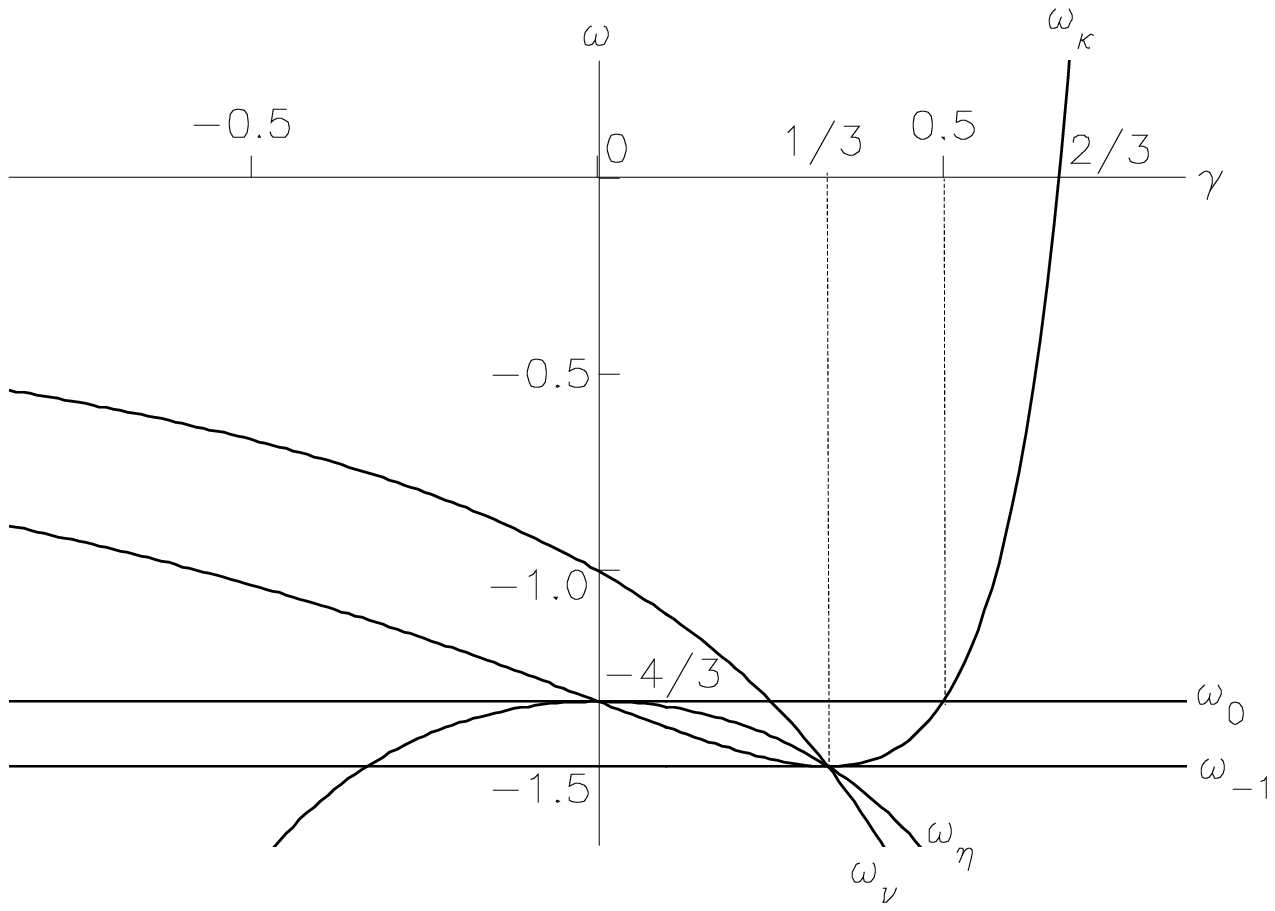}}$$
\end{center}
\end{figure}

\begin{figure}
\begin{center}
$${\epsfxsize=10.0 truecm \epsfbox{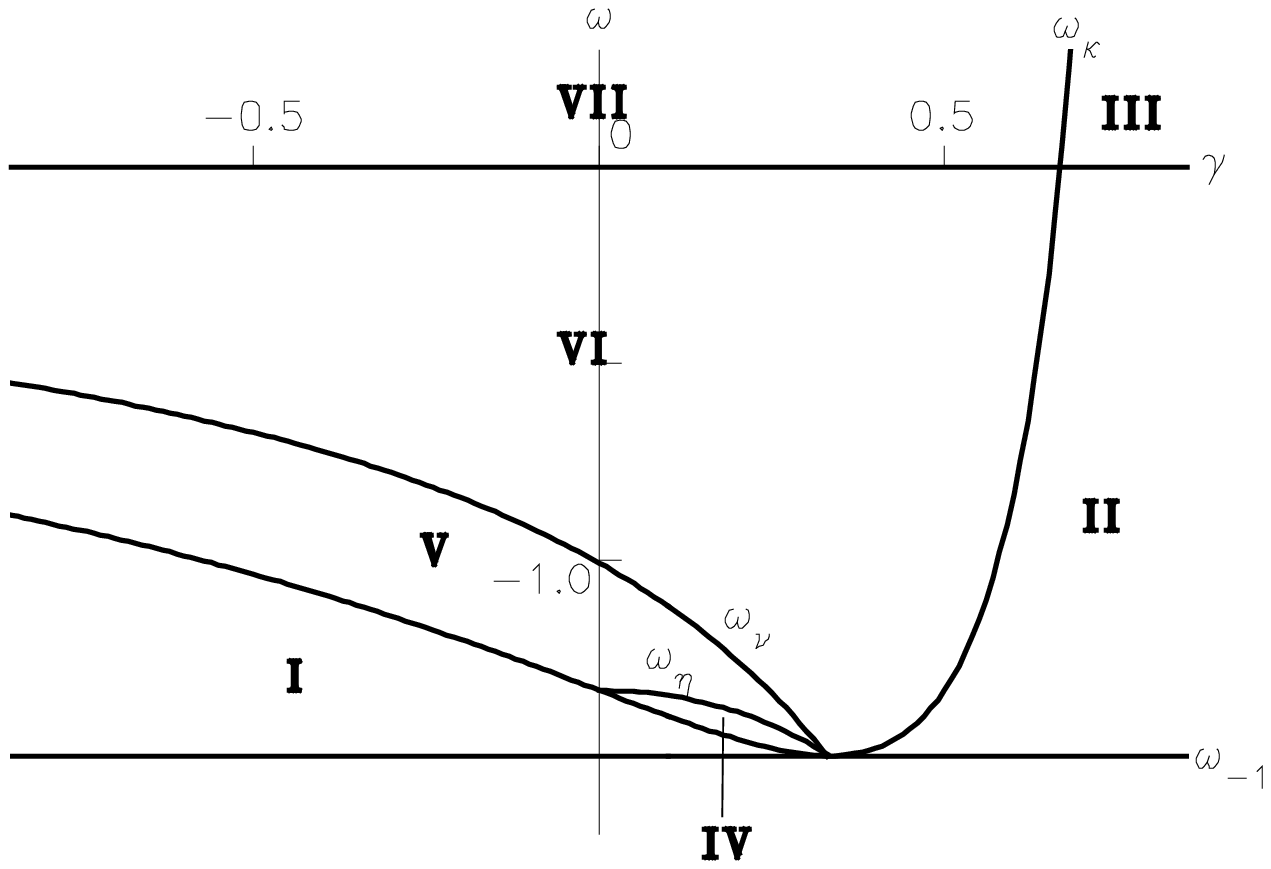}}$$
\end{center}
\end{figure}

\newpage

\begin{figure}
\begin{center}
$${\epsfxsize=10.0 truecm \epsfbox{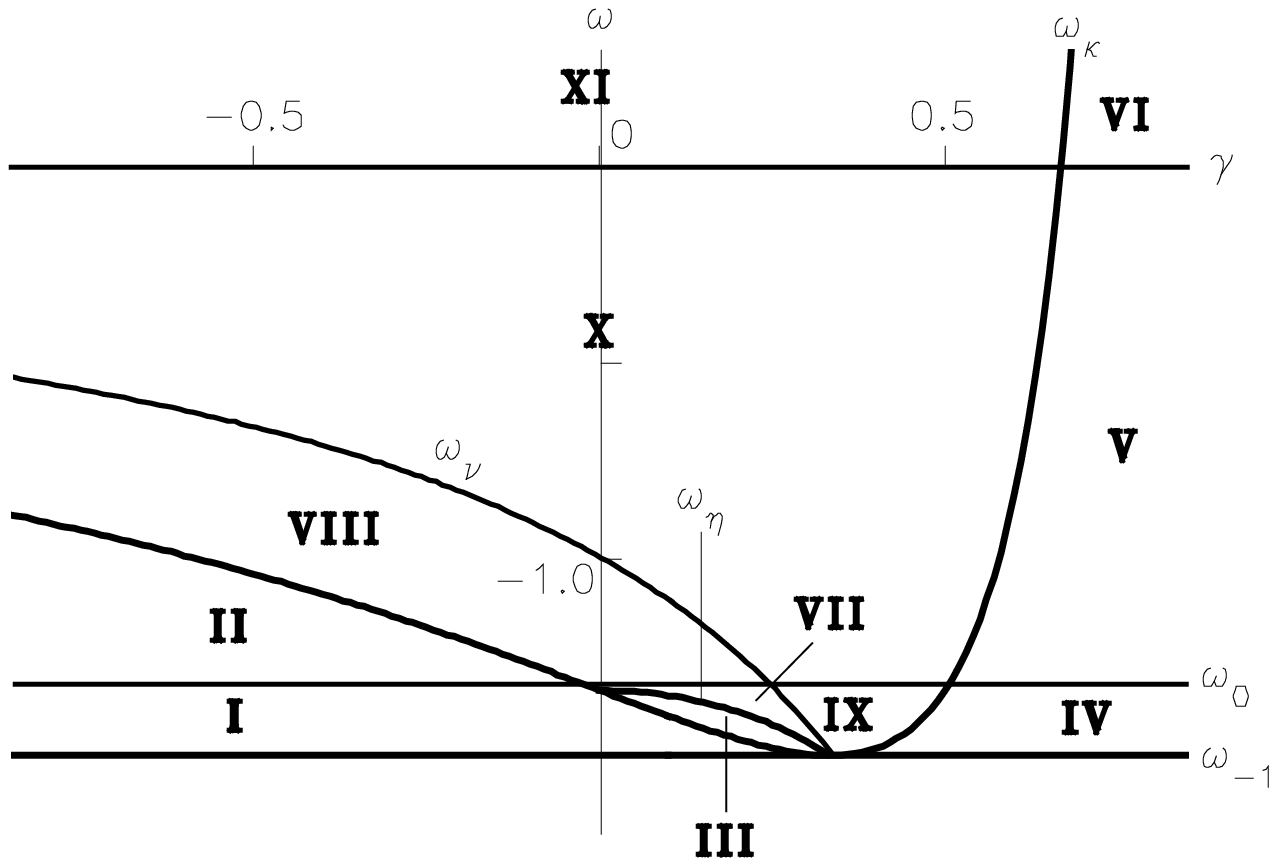}}$$
\end{center}
\end{figure}

\end{document}